# AUTOMATED CHARACTERIZATION OF SINGLE-PHOTON AVALANCHE PHOTODIODE

**AINA MARDHIYAH M. GHAZALI[1], AUDUN NYSTAD BUGGE[2], SEBASTIEN SAUGE[3] AND VADIM MAKAROV[2]**

[1]*Department of Science in Engineering, Faculty of Engineering, International Islamic University Malaysia, P.O Box 10, 50728 Kuala Lumpur, Malaysia.*
[2]*Department of Electronics and Telecommunications, Norwegian University of Science and Technology, NO-7491 Trondheim, Norway.*
[3]*School of Information and Communication Technology, Royal Institute of Technology (KTH), Electrum 229, SE-16440, Kista, Sweden.*

*ainamardhiyah71@gmail.com*

***ABSTRACT:*** We report an automated characterization of a single-photon detector based on commercial silicon avalanche photodiode (PerkinElmer C30902SH). The photodiode is characterized by I-V curves at different illumination levels (darkness, 10 pW and 10 µW), dark count rate and photon detection efficiency at different bias voltages. The automated characterization routine is implemented in C++ running on a Linux computer.

***ABSTRAK:*** Kami melaporkan pencirian pengesan foton tunggal secara automatik berdasarkan kepada diod foto runtuhan silikon (*silicon avalanche photodiode*) *(PerkinElmer C30902SH)* komersial. Pencirian diod foto adalah berdasarkan kepada plot arus-voltan (*I-V*) pada tahap pencahayaan yang berbeza (kelam - tanpa cahaya, 10pW, dan 10µW), kadar bacaan latar belakang, kecekapan pengesanan foton pada voltan picuan yang berbeza. Pengaturcaraan C++ digunakan di dalam rutin pencirian automatik melalui komputer dengan sistem pengendalian LINUX.

## 1. INTRODUCTION

Single-photon detectors (SPDs) are widely used for measuring extremely low light intensities. They have found diverse applications in laser ranging [1], astronomy [2], fluorescence detection [3], quantum optics, quantum information and quantum key distribution [4]. Nowadays, technologies for SPDs include photomultipliers, avalanche photodiodes (APDs), frequency up-conversion, visible-light photon counters, and several types of superconducting devices [5]. However, in real applications, APDs are often the most practical choice due to several advantages as compared to other photodetectors: small size, ruggedness, reliability, low sensitivity to magnetic fields and external disturbance in general, as well as lower cost [6, 7].

In order to detect single photons, the APD is operated in Geiger mode, and is also known as a single-photon avalanche diode (SPAD) [7]. In this mode the APD is biased above its breakdown voltage $V_{br}$. Single-photon sensitivity is achieved by exploiting the internal signal amplification, called avalanches, due to the process of impact ionization. In the Geiger mode, the electron-hole generation becomes self-sustaining, and one can register a macroscopic current flow due to a single incident photon. Currently, silicon

APDs are the most common choice for single-photon detection in the visible to near-infrared range up to ~1000 nm [4, 7].

There are several characteristics associated with the SPAD that need to be assessed prior to its use, for example its spectral range, dark count rate, dead time, photon detection efficiency, and timing jitter [7]. Often, characterization of many APD samples at different temperatures is required. Repeated characterization of the same sample may be useful in reliability studies and tests of APD's resilience against external factors, such as radiation and laser damage. In all these cases, a manual characterization would be impractical, and might also be less consistent. In this paper, we report a fully automated characterization of the I-V curves, dark count rate and photon detection efficiency of a PerkinElmer C30902SH commercial silicon avalanche photodiode. A custom testing setup has been built, and characterization programmed in C++ on a Linux platform.

## 2. EXPERIMENTAL SETUP

When the APD is reverse-biased above the breakdown voltage, an absorbed photon will trigger an avalanche event consisting of thousands of carriers. The current continues to flow until the avalanche is quenched by lowering the bias voltage to $V_{br}$ or below. In our single-photon detector, we use a simple and robust passively-quenched scheme (Fig. 1) [11–13]. The circuit consists of a high-voltage supply, 390 kΩ bias resistor, and a high-speed comparator for sensing the avalanche current. The avalanche current is initially sustained by charge stored in APD stray capacitance, however the voltage at the APD quickly drops and the avalanche self-quenches in about 1 ns. Then the capacitance is slowly charged via the bias resistor, and the detector recovers its sensitivity in ~1 μs. The APD is cooled with a thermo-electric cooler to a fixed temperature of –25 °C, in order to reduce its dark count rate [8].

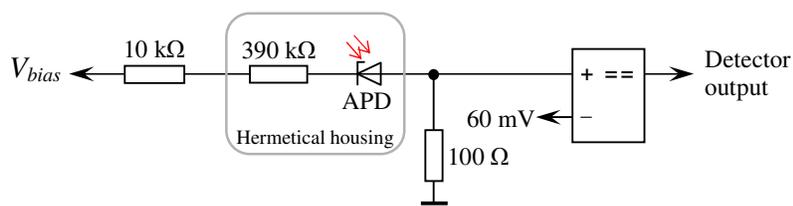

Fig. 1: Passively-quenched detector circuit.

Figure 2 depicts a schematic diagram of the characterization setup. Signametrics SMU 2055 multimeters were used to measure the bias voltage, average current flowing through the APD, and APD temperature. A Stanford Research Systems SR620 universal time interval counter was used to measure the frequency of photon counts registered at the detector. We used two signal lasers: one JDSU 54-00213 and one Sanyo DL-8141-002, both powered by a Highland Technology P400 signal generator. The lasers were coupled via single-mode optical fibers and 10/90 coupling ratio optical fiber coupler into one attenuated arm to the SPAD, and another arm to the power meter. The laser power was measured by a Newport 1830-C power meter using a Newport 818-SL-L photodetector head. An OZ Optics DA100 programmable attenuator was used to attenuate the laser light illumination to the desired intensity at the SPAD. The attenuation was calculated based on the power measured by the power meter and the manually determined splitting ratios of the coupler including fixed attenuation. All instruments were connected to the computer using RS-232 interface, except the SMU 2055 multimeters which were connected via USB interface.

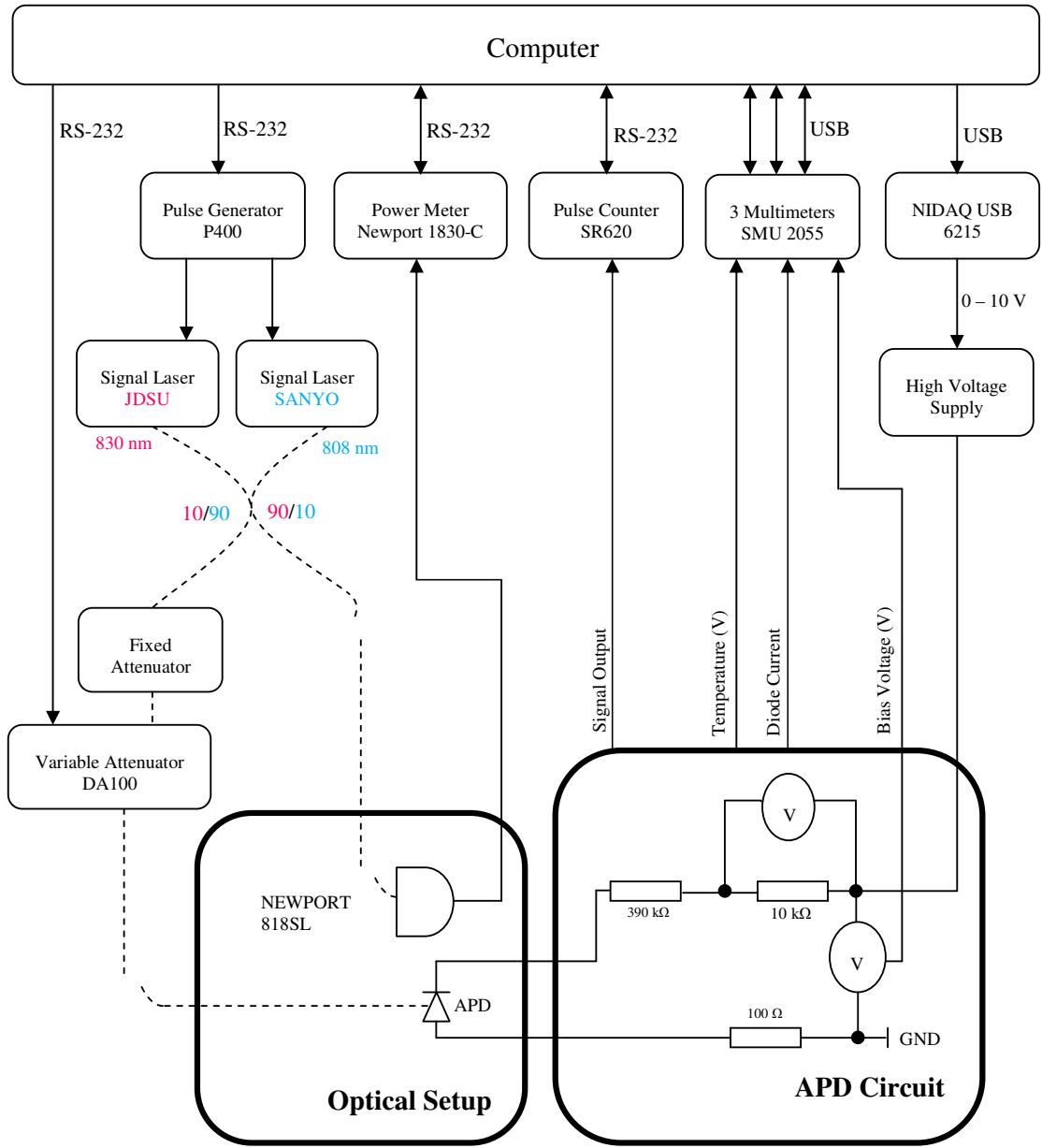

Fig. 2: APD characterization setup. Sufficient blackout measures were implemented to leave the APD in complete darkness when both lasers were unpowered.

In order to run complete characterization of SPAD as a function of bias voltage automatically, we should be able to

- a. control the power level of lasers and measure the power;
- b. set the bias voltage of the SPAD;
- c. measure the photon count rate registered by the detector;

d. measure the bias voltage applied;

e. measure the average current through the SPAD.

We implement the automated characterization routine using the object-oriented C++ programming language in a Linux environment. For data analysis, shell scripts and the open source numerical computation language/interpreter Octave were used.

## 3. SPAD CHARACTERIZATION PARAMETERS

In order to detect single photons, the SPAD must be operated in Geiger mode, i.e., it is biased above the breakdown voltage. *Overbias voltage* is defined as

$$V_{over} = V_{bias} - V_{br}, \tag{1}$$

where $V_{bias}$ is the bias voltage applied.

While operating in Geiger mode in darkness, spurious avalanches in the SPAD produce random pulses at a frequency known as the dark count rate (DC). These counts arise due to thermal carrier generation, band-to-band tunneling, and afterpulses (emission of trapped carriers from deep trap levels) [7, 12].

As $V_{bias}$ is increased, there is a sharp increase in the photodiode current, $I$ if the SPAD is under suitable illumination. A sharp bend in the I-V curve roughly coincides with $V_{br}$. The intensity of the incident light must be strong enough that it triggers many avalanches when the bias voltage increases past $V_{br}$, yet weak enough to not cause a significant amplified photocurrent below $V_{br}$.

Alternatively, $V_{br}$ can be determined from *threshold voltage* $V_{th}$. Threshold voltage is defined as the voltage at which photon counts start to appear at the detector output. $V_{th}$ is higher than $V_{br}$ by a fixed offset that depends on the comparator threshold setting, as the avalanche pulse must be of certain amplitude to be detected by the comparator. In our detector, this offset is 3.3 V ($V_{br}$ for calculating this offset value was determined manually by observing on an oscilloscope at which bias voltage small analog avalanche pulses begin to appear at the comparator input).

The photon detection efficiency (DE) is defined as the probability of detecting a photon incident on the detector. It depends on the diode's quantum efficiency and the probability for an electron-hole pair to trigger an avalanche [10]. The detection efficiency can be obtained through two distinct methods: by a calibrated light source [5] and by the correlated photon method [14]. We implemented the former. In this work the SPAD was illuminated with thousands photons per second at 830 nm. DE can be calculated as

$$\mathrm{DE} = (C - \mathrm{DC})/N, \tag{2}$$

where $C$ is the measured photon count rate, $N = P\lambda/hc$ is the calibrated incoming photon rate, $P$ is the continuous-wave (c.w.) optical power focused in the middle spot of the SPAD photosensitive area, $\lambda$ is the laser wavelength, $h$ is the Planck constant and $c$ is the speed of light. In our experiment, we restricted $P$ to 10 fW which corresponds to 41750 photons/s, in order to avoid having to take into account detector saturation effects [8]. In most photon-counting applications, a high value of DE is advantageous, however as we'll see later there is a tradeoff with increased DC.

## 4. MEASUREMENT METHODS AND ALGORITHMS

The entire SPAD characterization specified above is run from a C++ program on a computer. Each instrument was assigned with its own namespace which contained a class definition, and functions declaration for the instrument to be working accordingly. A main function was created to run the entire characterization program. For each measurement, one function was created to set the instruments as per measurement requirements.

For all measurements, the maximum overbias voltage was limited to 50 V (corresponding $V_{bias}$ known from a rough manual measurement), to avoid damage to the SPAD. The counting time was 10 s, except at $V_{over}$ = 15 V it was set to 100 s to reduce statistical uncertainty in data.

### 4.1 I-V Curves

The initial step in the automated characterization of the SPAD was to measure the breakdown voltage, $V_{br}$ from the I-V curve. The algorithm was as follows. As different bias voltages, $V_{bias}$ were applied across the SPAD, the potential drop over the 10 kΩ resistor $V_{10k}$ was measured by a SMU 2055 multimeter (see Fig. 2). Then, the current $I$ through the SPAD was calculated as

$$I = V_{10k}/10 \text{ k}\Omega. \qquad (3)$$

Next, the voltage across the APD $V_{APD}$ could be determined as

$$V_{APD} = V_{bias} - V_{R_{total}}, \qquad (4)$$

where $V_{R_{total}} = IR_{total}$, and $R_{total}$ = 390 kΩ (bias resistor) + 10 kΩ + 100 Ω = 400100 Ω. A plot of $I$ versus $V_{APD}$ was then created using these equations. All calculations and plots were done in Octave software.

I-V curves were obtained for three different illumination levels; zero, weak and medium illumination power. The purpose of the low power illumination was to detect the sharp increase in current which occurs at $V_{br}$. The purpose of the medium illumination power was to observe avalanche multiplication below $V_{br}$ and diode quantum efficiency at low values of $V_{bias}$ (when no multiplication occurs). The measurements were achieved using the following parameters.

 a. Zero illumination: Both lasers were off.

 b. Weak illumination: The APD was illuminated at 10 pW c.w. power at 830 nm.

 c. Medium illumination: The APD was illuminated at 10 µW c.w. power at 808 nm.

### 4.2 Threshold Voltage

Threshold voltage $V_{th}$ was obtained by measuring the dark count rate and the photon count rate (at 41750 photons/s) as a function of $V_{bias}$, using the SR620 counter. $V_{th}$ determination was divided into two parts: binary search, and linear search with 0.1 V increment. Binary search is needed because of the large range of possible $V_{bias}$. First, the search starts as a linear search in ±0.5 V range around a manually pre-calibrated value of $V_{th}$. At this point, if no counts were detected, or if counts were found at the lower bound of the linear search, the algorithm escapes to a binary search. The binary search will be conducted in the range between $V_{th}$ and the maximum $V_{bias}$ if the linear search produced no counts. It will be conducted in the range between 0 V and the initial $V_{th}$ if counts were registered at the lower bound of the linear search. The binary search seeks $V_{th}$ by repeatedly splitting the search range into half, until the range becomes smaller than 1 V. Then, a linear search will be executed over the remaining voltage range.

Unexpected conditions such as always zero counts or always non-zero counts over the full range of bias voltages are recognized by the program, and will be registered in the log file. The program is designed such that it will not crash or hang due to such conditions.

### 4.3 Dark Count Rate and Photon Detection Efficiency

Once $V_{br}$ is determined, the dark and photon count rates are measured as a function of overbias voltage, at $V_{over}$ = 1, 2, 3, 4, 5, 6, 7, 8, 10, 12, 14, 16, 18, 20, 25, 30, 35, 40, 45 V. The data is then used to calculate DE via Eq. 2.

## 5. RESULTS

Figure. 3 depicts the plots of $I$ versus $V_{APD}$ at three different illumination levels. At 10 pW illumination, the sharp bend indicating $V_{br} \approx 166.6$ V is clearly visible {Fig. 3(b)}. From Fig. 3(c), we calculate the quantum efficiency of 82 % at low bias voltage, when the APD has no internal gain.

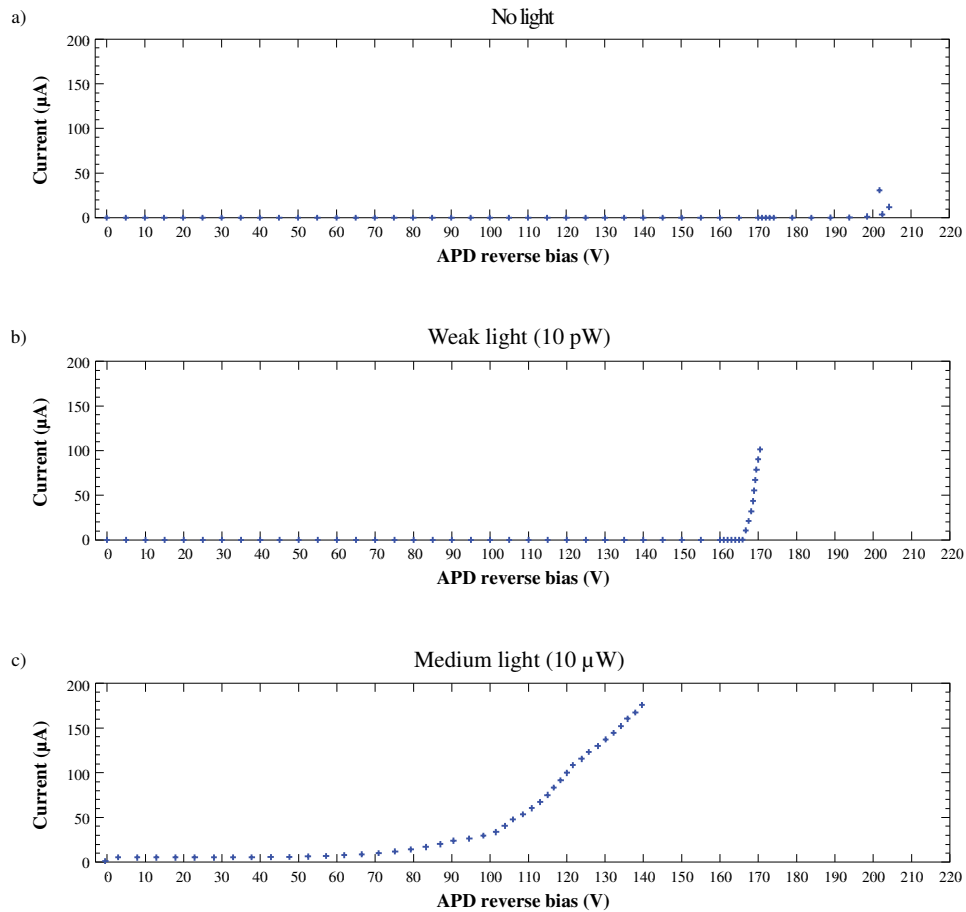

Fig. 3: I-V curves at (a) no illumination, (b) weak illumination (10 pW), and (c) medium illumination (10 µW).

The results of dark count rate and photon count rate measurements are shown in Fig. 4. The value of $V_{th} \approx 170$ V can be readily observed from both curves. DC increases with overvoltage. DE also increases, until it peaks at $V_{over} \approx 30$ V. The following decrease of the count rate is due to self-sustaining avalanches, which is a known characteristic of the passively-quenched scheme [8]. Based on the data of Fig. 4 and Eq. 2, DE is calculated, (Fig. 5). The highest photon count efficiency is $\approx 55\%$ at $V_{over} \approx 30$ V.

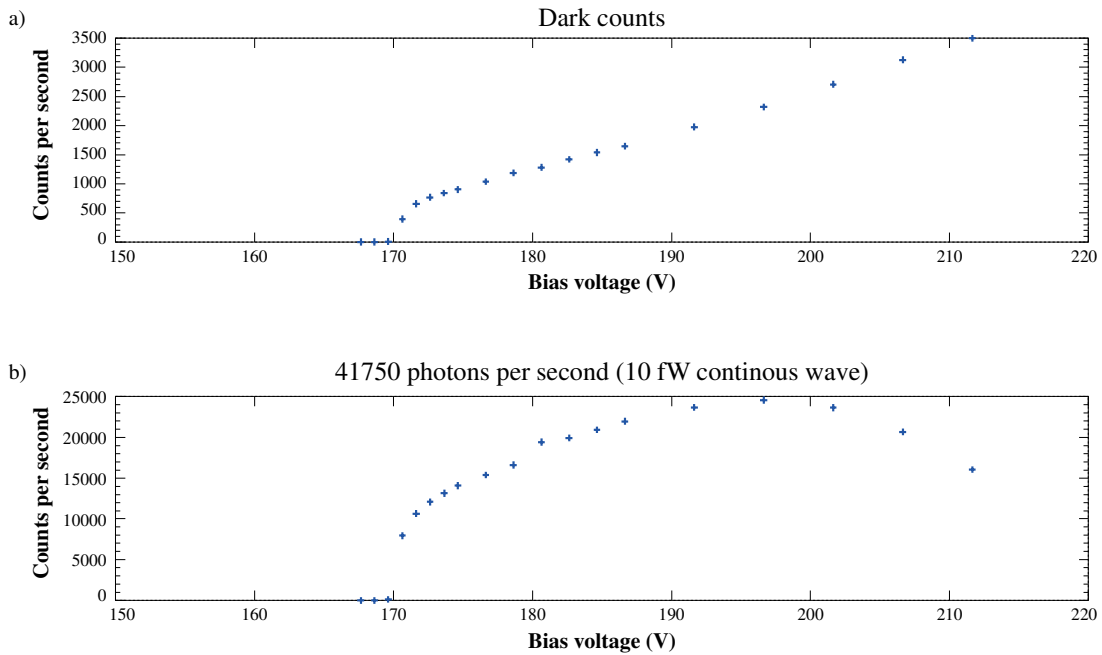

Fig. 4: Detector count rate vs. bias voltage at (a) no illumination (dark count rate), and (b) 10 fW c.w. illumination (41750 photons/s).

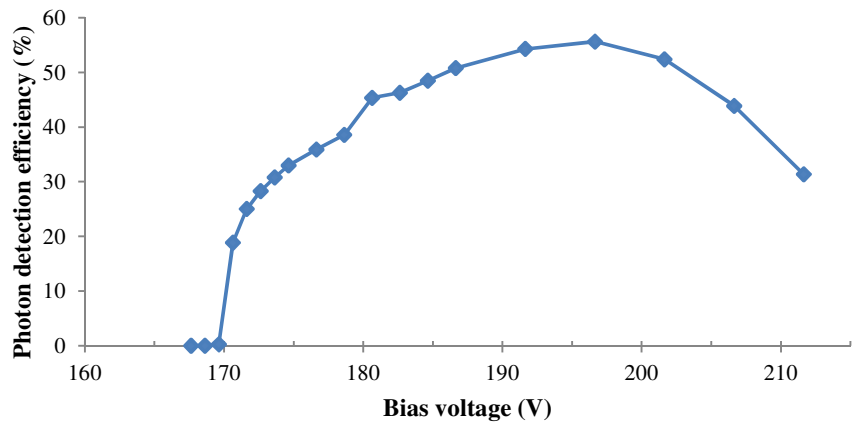

Fig. 5: Photon detection efficiency vs. bias voltage.

## 6. CONCLUSION

We demonstrated automated characterization of a SPAD. Using the developed algorithms, implemented in C++ and Octave on Linux platform, we were able to measure I-V curves, automatically determine breakdown voltage, measures dark count rate and photon detection efficiency. The developed characterization platform can be easily adapted and extended for different experimental needs.

## ACKNOWLEDGEMENTS

We thank Raihan Othman and Johannes Skaar for group leadership, Qin Liu and Lars Lydersen for help with programming, Christian Kurtsiefer for providing the detector electronics design. This work was supported by the Research Council of Norway (grant no. 180439/V30). A.M.M.G. acknowledges travel support from the International Shari'ah Research Academy for Islamic Finance (ISRA). S.S. acknowledges support from the ECOC'2004 foundation and the Research Council of Sweden (VR, grant no. 621-2007-4647).

## REFERENCES

[1] I. Prochàzka, K. Hamal, and B. Sopko, "Photodiode based detector package for centimeter satellite laser ranging", Proc. 7$^{th}$ Int. Workshop Laser Ranging Instrumentation, C. Veillet, OCA-CERGA Grasse, 1990, Ed., Matera, Italy, Oct. 2–8, 1989, pp. 219–221.

[2] N. S. Nightingale, "A new silicon avalanche photodiode photon counting detector module for astronomy", Exp. Astron. 1, 407–422 (1991).

[3] S. Cova, A. Longoni, A. Andreoni, and R. Cubeddu, " A semiconductor detector for measuring ultra-weak fluorescene decays with 70 ps FWHM resolution", IEEE J. Quantum El., QE-19, 630–634 (1983).

[4] N. Gisin, G. Ribordy, W. Tittel, H. Zbinden, "Quantum Cryptography", Rev. Mod. Phys. 74, 145–195 (2002).

[5] R. H. Hadfield, "Single-photon detectors for optical quantum information applications", Nat. Photonics 3, 696–705 (2009).

[6] Zappa *et al.*, "An integrated active-quenching circuit for single-photon avalanche diodes", IEEE Trans. Instrum. and Measurements 49, 6, 1167–1175 (2000).

[7] S. Cova, M. Ghioni, A. Lotito, I. Rech, and F. Zappa, "Evolution and prospects for single-photon avalanche diodes and quenching circuits", J. Mod. Opt. 51, 1267–1288 (2004).

[8] Y.-S Kim, Y.-C. Jeong, S. Sauge, V. Makarov, and Y.-H. Kim, "Ultra-low noise single-photon detector based on Si avalanche photodiode", Rev. Sci. Instrum. 82, 093110 (2011).

[9] F. Zappa *et al.,* "Integrated array of avalanche photodiode for single-photon counting", IEEE European Solid-State Device Research Conference (ESSDERC), 600–603 (1997).

[10] C. Niclass, M. Sergio, and E. Charbon, "A single photon avalanche diode array fabricated in deep-submicron CMOS technology", in Proc. DATE, 1–6 (2006).

[11] R. H. Haitz, "Model for the electrical behavior of a microplasma," J. Appl. Phys. 35, 1370–1376 (1964).

[12] R. H. Haitz, "Mechanisms contributing to the noise pulse rate of avalanche diodes," J. Appl. Phys. 36, 3123–3131 (1965).

[13] S. Cova, M. Ghioni, A. Lacaita, C. Samori, and F. Zappa, "Avalanche photodiodes and quenching circuits for single-photon detection," Appl. Opt. 35, 1956–1976 (1996).

[14] D. N. Klyshko, "Use of two-photon light for absolute calibration of photoelectric detectors," Sov. J. Quantum Electron. 10, 1112–1117 (1980).